\documentclass[twocolumn,epjc3]{svjour3}  

\usepackage{bm}
\usepackage{mathrsfs}
\usepackage{xcolor,color,graphicx,graphics}
\usepackage[all]{xy}
\usepackage{epsfig,subfigure}
\usepackage{latexsym,amssymb,amsmath,amsfonts} %\usepackage{amstex}
\usepackage[english]{babel} %, spanish, portuguese
\usepackage[OT1]{fontenc}
\usepackage[latin1]{inputenc}
\usepackage{makeidx}
\usepackage{hyperref}
\usepackage{color,graphicx,graphics,wrapfig,epsf}%,psfig
\usepackage{float}
\restylefloat{table}
\usepackage[colorinlistoftodos]{todonotes}
\usepackage{blindtext, xcolor}
\usepackage{verbatim}
\usepackage{ulem}
\definecolor{red}{rgb}{1,0,0}

\newcommand{\bi}{\begin{itemize}} 				\newcommand{\ei}{\end{itemize}}
\newcommand{\benu}{\begin{enumerate}} 		\newcommand{\enu}{\end{enumerate}}
\newcommand{\bd}{\begin{dinglist}{0}}     \newcommand{\ed}{\end{dinglist}}
\newcommand{\bfig}{\begin{figure}[htbp]}  \newcommand{\efig}{\end{figure}}
\newcommand{\bc}{\begin{center}} 				  \newcommand{\ec}{\end{center}}
\newcommand{\be}{\begin{equation}} 				\newcommand{\ee}{\end{equation}}
\newcommand{\bsub}{\begin{subequations}}  \newcommand{\esub}{\end{subequations}}
\newcommand{\ben}{\begin{eqnarray}} 			\newcommand{\een}{\end{eqnarray}}
\newcommand{\ba}[1]{\begin{array}{#1}} 		\newcommand{\ea}{\end{array}}
\newcommand{\bea}{\begin{equation}\begin{array}{rcl}}
\newcommand{\eea}{\end{array}\end{equation}}

\usepackage{soul}

\RequirePackage{fix-cm}
\smartqed  % flush right qed marks, e.g. at end of proof
\RequirePackage{graphicx}
\journalname{Eur. Phys. J. C}
\begin{document}

\title{Constraining Snyder and GUP models with low-mass stars%\thanksref{t1}
}


\author{Anna Pacho\l\thanksref{e1,addr1}
        \and
       Aneta Wojnar\thanksref{e2,addr2} %etc.
}

\thankstext{e1}{e-mail: anna.pachol@usn.no}
\thankstext{e2}{e-mail: awojnar@ucm.es}


\institute{ Department of Microsystems, University of South-Eastern Norway, Campus Vestfold,\\ Raveien 215, 3184 Borre, Norway \label{addr1}
           \and
           Departamento de F\'isica Te\'orica \& IPARCOS, Universidad Complutense de Madrid, E-28040, 
Madrid, Spain \label{addr2}
}

\date{Received: date / Accepted: date}

\maketitle

\begin{abstract}
We investigate the application of an equation of state that incorporates corrections derived from the Snyder model (and the Generalized Uncertainty Principle) to describe the behaviour of matter in a low-mass star. Remarkably, the resulting equations exhibit striking similarities to those arising from modified Einstein gravity theories. By modeling matter with realistic considerations, we are able to more effectively constrain the theory parameters, surpassing the limitations of existing astrophysical bounds. The bound we obtain is { $\beta_0 \leq 4.5 \times 10^{47}$}. We underline the significance of realistic matter modeling in order to enhance our understanding of effects arising in quantum gravity phenomenology and implications of quantum gravitational corrections in astrophysical systems.
\end{abstract}

\section{Introduction}
Quantum gravity research and the search for its observable effects have been drawing ever increasing interest.  To reconcile the principles of quantum mechanics and general relativity we must challenge the foundational concepts of classical space-time and continuous symmetries. {Quantum gravity suggests that, at the Planck scale, the known structure of space-time should be modified}, at the same time affecting the quantum-mechanical phase-space.
The first example of deformed (non-commutative) coordinate space as a background for the unification of gravity and quantum field theory was introduced in \cite{Doplicher:1994tu,Doplicher:1994zv}. Such feature also appeared in String Theory, with a constant background field, where coordinates on the space-time manifolds at the
end of strings (D-branes) do not commute \cite{Seiberg:1999vs}. Additionally, the minimal length or arguments that the space-time can no longer be continuous appear in various quantum gravity approaches \cite{Gross:1987ar,Rovelli:1994ge,Ambjorn:2005db,Lauscher:2005qz,Horava:2009if}.
%such as Loop quantum gravity
%(Nucl.Phys.B 442 (1995), 593-622), Causal Dynamical Triangulations Phys.Rev.Lett. 95 (2005), 171301,
%Asymptotically Safe Quantum Gravity JHEP 10 (2005), 050, Horava-Lifshitz Gravity Phys.Rev.Lett. 102 (2009), 161301, Nucl.Phys.B 303 (1988), 407-454).}
Such modifications of space-time should have implications on physical phenomena such as gravitational and cosmological effects. This way we can try to model the quantum gravitational corrections in an effective description without full knowledge of quantum gravity theory itself. In the non-commutative (NC) geometry approach, it is assumed that the quantization process of general relativity, must include the quantization of space-time, and the space-time coordinates become non-commutative. This, in turn, affects the Heisenberg uncertainty relation, through the modifications of the phase space. Physical effects, relying on such modified uncertainty relation, can often be expressed in the form of quantum gravitational corrections to classical solutions and may provide physical predictions, guiding us in the choice of the correct quantum gravity model, when compared against measurements.
It should be noted that various approaches to quantum gravity may introduce such quantum gravitational corrections, resulting in an appearance of high curvature terms in gravitational Lagrangians \cite{Donoghue:1993eb}. Consequently leading to the modified Einstein equations, which we will refer to as "modified Einstein gravity" in this text.

The introduction of the generalized Heisenberg uncertainty principle (GUP) \cite{Maggiore:1993rv,Maggiore:1993zu,Kempf:1994su,Chang:2001bm} attracted a lot of attention due to its potential for measurable outcomes, see e.g. \cite{Wang:2010ct,Ali:2011ap,Ali:2013ii,Moussa:2015bsa,Pachol:2023tqa,Bernaldez:2022muh,Tunacao:2022ffq,Mathew:2017drw,Tamburini:2021inp,Das:2021lrb}. GUPs have been quite fruitful in investigating measurable effects of quantum gravitational nature, irrespective of the specific model of NC space-time. For instance, extensive research utilizing the GUP approach has provided bounds on the minimal length, although these bounds often deviate from the expected order of the Planck length ($L_P\sim \sqrt{\frac{\hbar G}{c^3}}$). Nonetheless, these investigations have yielded many insights into the subject \cite{Scardigli:1999jh,Chang:2001kn,Brau:2006ca,Das:2008kaa,Scardigli:2016pjs,Harikumar:2017suv,barca2019semiclassical,Segreto:2022clx,Campbell:2023dwm}, see also \cite{Bosso:2023aht} for a recent review and more references on the topic.

NC space-times can be considered as a natural framework for GUP theories since they lead to deformations of the Heisenberg uncertainty principle leading to its generalizations. However the minimal length may not necessarily appear if one bases only on the modifications of the phase space, see e.g. \cite{Segreto:2022clx,Bishop:2019yft,Bishop:2022vzr,Bosso:2023sxr}.

One of the quantum space-time models, which may be associated with GUP framework is the Snyder space \cite{snyder71quantized}. It was proposed already in 1947 and is known as the first example of Lorentz-covariant NC space-time, admitting a fundamental length scale.
  The non-commutativity of the Snyder {space}
   is encoded in the relation between the {spacial} coordinates $\hat{x}_{i}$ as
  %and is proportional to the Lorentz rotations $M_{ij}$
$\lbrack \hat{x}_i,\hat{x}_j]=i\hbar\beta M_{ij}$, {where indices $i,j=1,2,3$.} \footnote{Here we consider the non-relativistic Euclidean case {to be in accordance with GUP approach}. {However in the context of NC geometry the full Snyder model describes quantum space-time and is defined by $\lbrack \hat{x}_{\mu},\hat{x}_{\nu}]=i\hbar\beta M_{\mu\nu}, \quad\mu,\nu=0,1,2,3 $ where $\hat{x}_{\mu}$ are NC space-time coordinates and $M_{\mu\nu}$ are the generators of the Lorentz algebra which is the symmetry of this NC space-time, .}}.
Parameter $\beta $ is the deformation parameter, of dimension $\left[\frac{L^2}{\hbar^2}\right]=[p^{-2}]$, that sets the scale of non-commutativity (as $L$ length is usually associated with the Planck length $L_{p}$). 
The phase space associated with the Snyder model, involves modified commutation relations between coordinates and momenta and, up to the linear order in the non-commutativity parameter $\beta$, can be expressed as follows \cite{Meljanac:2021iyk}:
\begin{eqnarray}
\left[ p_i,\hat{x}_j\right] &=&-i\hbar\delta_{ij}\left( 1+\beta \left( \chi-\frac{1}{2}\right) p_kp_k\right)\nonumber\\
&-&2i\hbar\chi\beta p_ip_j+O(\beta ^{2}). \label{gen_real_p-x}
\end{eqnarray}
This form of deformed phase space bases on the 'general realization' of the Snyder model, proposed in \cite{Battisti:2010sr} and allows us to explore if any measurable effects favour a particular {realization} (the choice of $\chi$)\footnote{We point out that there is a distinction between the (Heisenberg) realization and the Heisenberg representation and the Hilbert space representation, see e.g \cite{Borowiec:2010yw} for details. One can show that different realizations of NC space-times, as well as different bases of quantum groups of symmetries, may lead to different physical effects, see e.g. \cite{Borowiec:2009ty}, \cite{Meljanac:2012pv}.}.
The original Snyder realization \cite{snyder71quantized}
is recovered for $\chi = 1/2$.
%, giving the usually considered phase space relations:
%\begin{equation}
%\left[ p_{\mu},\hat{x}_{\nu}\right] =-i\hbar(\eta _{\mu\nu}+\beta p_{\nu}p_{\mu}).%\label{c12}%
%\end{equation}
For $\chi=0$, one obtains the type of realization which can be linked to \cite{Maggiore:1993rv}, \cite{Maggiore:1993zu}. 
These two choices of realizations of phase spaces have been widely investigated in the context of GUP theories \cite{Maggiore:1993rv,Moussa:2015bsa,Ali:2011ap,Ali:2013ii,Bernaldez:2022muh,Tunacao:2022ffq,Mathew:2017drw,Wang:2010ct,Maggiore:1993zu,Kempf:1994su,bishop2020modified}. 

In this paper, we are interested in using the most general one-parameter family of deformed phase spaces (\ref{gen_real_p-x}) corresponding to the Snyder model and investigate the possible measurable effects associated with this type of noncommutativity {and their dependence on the choice of realization}. 

The form of the generalized phase-space (\ref{gen_real_p-x}) implies {one of the possible definitions of }  of the inner product\footnote{{Different momentum space representations of $\hat{x}_i$ and $p_{j}$ will lead to different inner products, here we have chosen the following:
$\hat{x}_{i}\phi(p)=i\hbar\left( \left( 1+\beta \left( \chi-\frac{1}{2}\right) p^{k}p_k\right) \frac{\partial }{\partial p_{i}}+2\chi\beta p_{i}p_{j}\frac{\partial }{\partial
p_{j}}+\gamma p_{i}\right) \phi(p)$ and $p_{i}\phi(p)=p_i\phi(p).$}} in the momentum space \cite{Pachol:2023tqa} (cf. \cite{Kempf:1994su}, \cite{Chang:2001bm}):
\begin{equation}\label{inn_prod_measure}
    \langle\psi,\phi\rangle = 
     \int \frac{d^3{p}}{(1+\omega p^2)^{\alpha}}\psi^*({p}) \phi({p})
\end{equation}
where $\alpha=\frac{(5\chi-\frac{1}{2})}{(3\chi-\frac{1}{2})}$ is dimensionless and $\omega=\beta(3\chi-\frac{1}{2})$. 
It is worth to point out that the values of $\alpha$ and $\omega$ are closely related with the choice of the realization parameter $\chi$. 
%Since we are interested in $D=3$ dimensional case, we have the following relation:
%\begin{equation}\label{par}
%\alpha\omega = \beta(5\chi-\frac{1}{2}).
%\end{equation}
Since the deformed Heisenberg algebra inducing a measure in momentum space depends on the representation, hence we must consider the full phase space measure via the Liouville theorem instead (to find the Liouville measure one computes the determinant of the symplectic form of the phase space), see eq. (9) in \cite{Chang:2001bm}.
For $D=3$ the phase space volume element, corresponding to (\ref{gen_real_p-x}), considered up to linear terms in $\beta$, will be modified as follows:
\begin{equation}\label{measure}
     \frac{d^3xd^3{p}}{1+\Omega p^2},\qquad \mbox{where} \quad\Omega =\beta
\left( 4\chi -\frac{3}{2}\right)
\end{equation}
 (for details see the Appendix and \cite{Chang:2001bm}).
Due to the modification of the phase space, the thermodynamical relations and hence physical effects are changed with respect to the standard case. In what follows, we will be interested in the Fermi gas equation of state which is commonly used in stellar and substellar physics. It takes various forms, depending on physics we want to incorporate into it; therefore, we consider the polytropic equations of state \cite{horedt2004polytropes} in the temperature $T\rightarrow0$ limit. One can also introduce the finite temperature corrections \cite{auddy2016analytic}, allowing to study a more realistic models of relativistic and non-relativistic stars \cite{glendenning2012compact,hansen2012stellar}, as well as substellar objects \cite{Burrows:1992fg,Seager:2007ix}.

 Fermi equation of state resulting from phase space deformations and the measure (\ref{measure}) associated with the Snyder NC space can be obtained by considering the partition function in the {grand-canonical ensemble}, which in the spherical symmetric case can be written as \cite{Pachol:2023tqa}: 
\begin{equation}\label{partition1}
     \mathrm{ln}Z =  \frac{V}{(2\pi \hbar)^3}\frac{g}{a}\int \mathrm{ln}\left[1+az e^{-E/k_BT}\right] \frac{4\pi p^2 dp}{{(1+\Omega p^2)}} \ ,
\end{equation}
where $T$ is the temperature, $k_B$ Boltzmann constant, $z=e^{\mu/k_BT}$, while $\mu$ is the chemical potential and $a=1$ ($a=-1$) if the particles are fermions (bosons) and $g$ is a spin of a particle, $V:=\int d^3x$ is the volume of the cell (of the configuration space).  By having the appropriate partition function we can find the thermodynamic variables such as pressure, number of particles, and internal energy by following the usual relations. For example, the pressure $P=k_B T\frac{\partial}{\partial V} \mathrm{ln}Z$, with $a=1$ for fermions and $g=2$ for electrons, becomes \cite{Pachol:2023tqa}:
\begin{equation}\label{pressuregen}
      P=  \frac{1}{\pi^2 \hbar^3}\int \frac{1}{3}p^3\, _{2}F_{1}\left(\frac{3}{2},{1},\frac{5}{2},{-p^2\Omega}\right) f(E) \frac{c^2 p}{E}dp,
\end{equation}
 where $f(E)$ is the Fermi-Dirac distribution $f(E)=\left(1+z e^{-E/k_BT}\right)^{-1}$
and $_{2}F_{1}$ is the hypergeometric function. From the properties of the hypergeometric function we can expand it for the case when {$|\Omega p^2| << 1$}, and take into account only the first two terms (as we consider the NC deformation only up to linear terms in $\beta$)\footnote{From the physical point of view, we are interested in temperatures in stellar objects which are below the Planck temperature.}: 
\begin{equation}
      P=  \frac{1}{\pi^2 \hbar^3}\int \frac{p^3}{3} \left( 
      \sum_{k=0}^{\infty} \frac{\left(\frac{3}{2}\right)_k(-{\Omega p^2})^k}{ \left(\frac{5}{2}\right)_k k!}
      \right) f(E) \frac{c^2 p}{E}dp.
\end{equation}
 Using the expression for energy of the non-relativistic electrons $E\approx\frac{p^2}{2m_e}$ and considering the case when $T\rightarrow0$, the integration of (\ref{pressuregen}) is taken up to the Fermi energy $E_F$ resulting in \cite{Pachol:2023tqa}:
\begin{equation}\label{pres3}
      P_{T\rightarrow0}=  \frac{2}{5}vE_F^\frac{5}{2} \left(1 -\frac{3{\Omega}}{7} (2m_e) E_F\right) ,
\end{equation}
where we have defined $v=\frac{(2m_e)^\frac{2}{3}}{3\pi^2 \hbar^3}$. We note that the pressure $P_{T\rightarrow0}$ becomes smaller when $\Omega>0$ while it increases for $\Omega<0$, depending on the choice of realization $\chi$ parametrizing the Snyder model. 

We can further rewrite (\ref{pres3}) in a more familiar form. To do that, let us use the definition of the measure of electron degeneracy, where $u=(3 \pi^{2} \hbar^{3} N_A)^\frac{2}{3}/2m_e$: 
\begin{equation}\label{degeneracy}
\psi=\frac{k_{B} T}{E_{F}}=\frac{2 m_{e} k_{B} T}{\left(3 \pi^{2} \hbar^{3}\right)^{2 / 3}}\left[\frac{\mu_{e}}{\rho N_{A}}\right]^{2 / 3}\equiv u^{-1}  k_{B} T \left[\frac{\mu_{e}}{\rho }\right]^{2 / 3}
\end{equation}
and rewrite (\ref{pres3}) as:
\begin{equation}\label{pres4}
    P_{T\rightarrow0} =\frac{2}{5}vu^\frac{5}{2}\left(\frac{\rho}{\mu_e}\right)^\frac{5}{3}\left[
    1-\frac{3u}{7} {\Omega} (2m_e)\left(\frac{\rho}{\mu_e}\right)^\frac{2}{3} 
    \right].
\end{equation}
Therefore, the modification introduced by NC deformation has a polytropic form.
\section{Lane-Emden equation in Snyder model}
As briefly recalled in the previous section, our previous work \cite{Pachol:2023tqa} demonstrated that Fermi equation of state in the limit $T\rightarrow0$ becomes the polytropic one (\ref{pres4}) with the additional term arising from the non-commutativity of space-time and deformation of the phase-space and is related with GUP. This term is parametrized by $\chi$ (now included in {$\Omega$}, instead of $\alpha\omega$ considered in \cite{Pachol:2023tqa}) and is governed by the choice of the realization of the Snyder model.
The thinking behind NC geometry approach is that non-commutativity should arise due to quantum gravity effects and in investigating phenomenological effects of these we can study quantum gravitational corrections to classical solutions. This is in accordance with GUP approach, where
the existence of a minimum measurable length and the related generalized uncertainty principle, influence all quantum Hamiltonians, hence predicting quantum gravity corrections to various quantum phenomena \cite{Das:2008kaa}.

In the following, we want to investigate the NC effects (appearing in the Fermi EoS) in a semi-classical regime describing a spherical symmetric ball made of Fermi gas.
%Since the corrections arising from the modifications given by the Snyder model are assumed to be very small (order of Planck length) [in relation to what? - we are working in a regime with strong gravity and quantum interactions,  or are we saying quantum gravitational corrections are small with respect to classical gravity?] 
Hence we consider the usual Poisson equation for the gravitational potential $\phi$
\begin{equation}\label{poisson}
    \nabla^2\phi =  4\pi G \rho
\end{equation}
where $G$ is the Newton constant and $\rho$  is an energy density, assuming
 $\rho=\rho(r)$ since we investigate the spherical-symmetric case. Because of that, all quantities appearing further in the paper will be only $r$-coordinate dependent.
 Additionally, the usual non-relativistic hydrostatic equilibrium is given by:
\begin{equation}\label{hydro}
    \frac{d\phi}{dr}=-\rho^{-1}\frac{dP}{dr},
\end{equation}
while mass is
\begin{equation}\label{mass}
     M=\int 4\pi'\tilde r^2 \rho(\tilde r) d\tilde r.
\end{equation}
We will consider these classical equations in the non-relativistic limit in this form, that is, we will assume that the corrections arising from the (unknown) quantum gravity theory are higher than the second order in velocities, therefore they do not appear in the Poisson equation (\ref{poisson}). However, we should keep in mind that the effective relativistic theories (modified Einstein gravity proposals) coming from quantum gravity can include terms which survive in the non-relativistic limit \cite{Olmo:2019flu,Toniato:2019rrd}. 
%\textcolor{red}{Comment: expand on that these are all usual (classical) equations and add motivation.}
Therefore, the only modification (interpreted as quantum-gravitational correction) appears in the pressure (\ref{pres4}), which is then used in the non-relativistic hydrostatic equilibrium equation (\ref{hydro}). We rewrite our equation of state (\ref{pres4}) in a simplified notation, as:
\begin{equation}\label{pres5}
    P_{T\to 0} =K_1{\rho}^\frac{5}{3}\left[
    1-\varepsilon{\rho}^\frac{2}{3}
    \right]
\end{equation}
where $K_1= \frac{2}{5}vu^\frac{5}{2}\mu_e^{-\frac{5}{3}}$ and the parameter 
\begin{equation*}
    \varepsilon=\frac{3}{7}(\frac{3\pi^2\hbar^3N_A}{\mu_e})^{\frac{2}{3}}{\Omega}=4.47878\times10^{-52}{\Omega}.
\end{equation*}
Now let us introduce the Lane-Emden formalism \cite{horedt2004polytropes}, that is, the Lane-Emden dimensionless quantities:
\begin{equation}\label{dimensionless variables}
r=r_c \xi,\;\;\;
\rho= \rho_c  [\theta(\xi)]^n ,\;\;\; P=p_c [\theta(\xi)]^{n+1},
\end{equation}
where $\rho_c$ and $p_c$ are the star's central density and pressure, respectively, and $\xi$ is a dimensionless radial coordinate while $\theta$ can be treated as dimensionless temperature. The parameter
$r_c$ is defined via the expression
 \begin{equation}\label{Eq: rc}
 r_c^2= \frac{2 (n+1) p_c }{4\pi G \rho_c^2  } = \frac{2(n+1)K \rho_c^{1/n-1}}{4\pi G} \ .
 \end{equation}
 This allows us to rewrite the Poisson equation (\ref{poisson}) in the following way.
 With the use of (\ref{hydro}) and the modified (\ref{pres5}) we get the {\it modified} Lane-Emden equation with $n=3/2$ 
\begin{equation}\label{Eq. Lane-Emden}
\frac{d}{d\xi}\left\lbrace  \xi^2  \, \frac{d\theta}{d\xi} \left[ 1 -\epsilon\theta   \right] \right\rbrace= - \xi^2 \theta^\frac{3}{2} \ ,
\end{equation}
where $\epsilon=\frac{7}{5}\varepsilon\rho_c^\frac{2}{3}=6.2703\times10^{-52}\rho_c^\frac{2}{3}{\Omega}$ which includes corrections arising from the Snyder model. The boundary conditions are given by $\theta(0)=1$ and $\theta'(0)=0$. %non-commutativity. \aw{napisz boundary conditions}

Note that the above equation can be considered {independently} as a modified Lane-Emden equation, arising from some {model of} modified Einstein gravity (for more details, see review \cite{Olmo:2019flu} and \cite{Wojnar:2022txk}). Indeed, (\ref{Eq. Lane-Emden}) corresponds to a modified Poisson equation
\begin{equation}\label{poisson2}
   \nabla^2\phi =  4\pi G \rho - \tilde\epsilon \nabla^2 \rho^\frac{4}{3}.
\end{equation}
On the other hand, if we consider {\it unmodified} polytropic equation of state
 $ P=K_1{\rho}^\frac{5}{3}$ and use (\ref{hydro}) and apply them to (\ref{poisson2}) we will obtain (\ref{Eq. Lane-Emden}).
 It means (\ref{pres4}) is taken in the limit of $\beta\to 0$ (of the usual commutative space-time) with the modified Poisson equation. The parameter $\tilde\epsilon$ appearing in the modified Poisson equation is re-scaled as $\tilde\epsilon=\frac{7}{4} K_1 \epsilon $.

Therefore, if one day we observe any such an effect in widely understood stellar astrophysics, we should remember that the effects of NC space-time as an approach to quantum gravity may be indistinguishable from effects of modified Einstein gravity approach. On the other hand, one should expect that a future quantum gravity theory will reduce to a modified Einstein gravity, such that any modification with respect to Newtonian physics appearing in, for instance, stellar equations, should not surprise us when we deal with approaches to quantum gravity models.

Regardless of the origin, NC geometry or modified Einstein gravity, the modified Lane-Emden equation (\ref{Eq. Lane-Emden}) can be used to rewrite the mass function, radius, central density $\rho_c$, and the temperature profile $T$ in terms of its solutions (note that we consider the case when $n=3/2$):
\begin{eqnarray}
M &=& 4 \pi \rho_c r_c^3 \omega,\label{Eq: star mass}\\
R&=& \gamma  \left( \frac{K}{G}\right)  M^{\frac{1-n}{3-n}} \ , \label{Eq: star radius}\\
\rho_c &=&\delta\left( \frac{3M}{4\pi R^3}\right) \ , \label{Eq: central dens}\\
T&=&\frac{K \mu}{N_Ak_B} \rho_c^{2/3}\theta \ , \label{Eq: temperature}
\end{eqnarray}
where $k_B$ is Boltzmann's constant, $N_A$ the Avogadro number and $\mu$ the mean molecular weight. The central temperature is defined as $ T_c = \frac{K \mu}{N_Ak_B} \rho_c^{2/3}$. The constants {$\Omega$}, $\gamma$, and $\delta$ are defined with respect to the solutions of the generalized Lane-Emden equation (\ref{Eq. Lane-Emden}), taken at the star's surface $\xi_R$ (given by $\theta(\xi_R)=0$, where $\xi_R$ is a dimensionless star's radius)
\begin{eqnarray}\label{parametry}
\omega&=& \left[-\xi^2  \, \frac{d\theta}{d\xi} \right]_{\xi=\xi_R},  \\
\gamma&=&(4\pi)^{\frac{1}{n-3}} (n+1)^{\frac{n}{3-n}} \, \omega_n^{\frac{n-1}{3-n}}  \xi_R,\\
\delta &=&  = -\frac{\xi_R}{3 \frac{d\theta}{d\xi}\Big|_{\xi=\xi_R}}.\label{delta}
\end{eqnarray}

  \begin{figure}[t]
\centering
\includegraphics[scale=0.5]{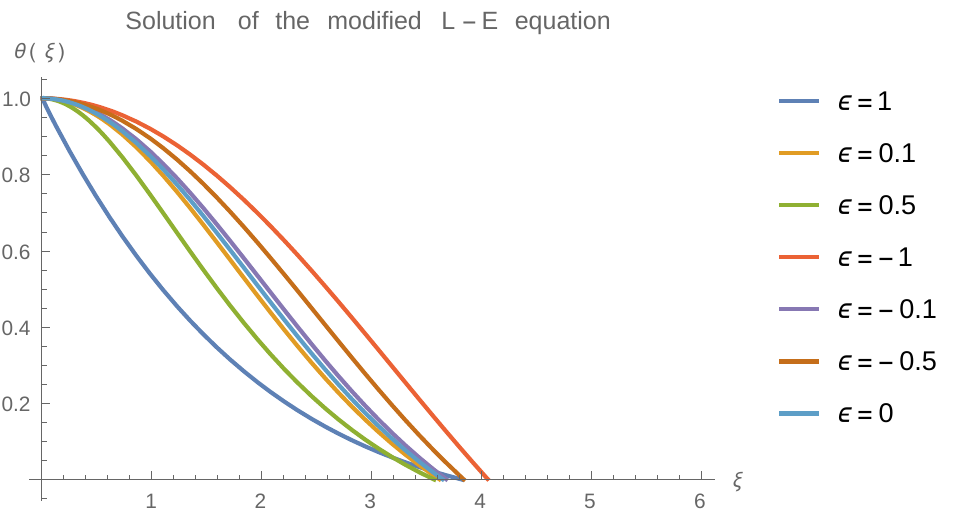}
\caption{Solutions of the Lane-Emden equation with $n=3/2$ in Snyder model for a few values of the parameter $\epsilon$.}
\label{LEplot}
\end{figure}

Generally, analytic solutions to the modified Lane-Emden equation are difficult to get, or impossible (even in Newtonian case there is no exact solution for $n=3/2$), and hence one can solve this equation numerically.
However, we can usually find an approximate solution at the center of the star $\theta(\xi\approx0)$, which in our case has the following form:
\begin{equation}\label{Eq: Theta}
\theta(\xi \approx 0)= 1- \frac{\xi^2}{6(1-\epsilon)} \sim \mathrm{exp}\left( -\frac{\xi^2}{6(1-\epsilon)}\right).
\end{equation}
Note that it depends on the parameter $\epsilon$, and it is singular for $\epsilon=1$ although the general numerical solution (see Fig. \ref{LEplot}) is not for this particular value.

Nevertheless, with the numerical solutions of (\ref{Eq. Lane-Emden}) we can compare the physical properties of a fully convective star from this toy-model. One can get them by applying the solutions to (\ref{parametry}-\ref{delta}) and utilizing the formulas (\ref{Eq: star mass}-\ref{Eq: temperature}). The results are given in the Table \ref{tab}.

\begin{table}[]
\begin{center}
\begin{tabular}{||c | c | c | c | c||}
 \hline
 $\epsilon$ & $M/M^{N}$ & $R/R^{N}$ & $\rho_c/\rho_c^{N}$ & $T_c/T_c^{N}$ \\ [0.5ex] 
 \hline\hline
$ -1$ & 1.95 & 1.39 & 0.71 & 0.79 \\
 $-0.5$ & 0.99 &1.08 & 1.16 & 1.11 \\
$-0.1$ & 1.08 & 1.04 & 0.95 & 0.97 \\
\hline
$0$ (Newt) & 1 & 1 & 1 & 1\\
\hline
$0.1$ & 0.92 & 0.96 &1.06 & 1.04 \\
$0.5$ &0.64 & 0.84 &1.46 & 1.29 \\
$1$& 0.4 & 0.77 & 2.86 &2.01\\ 
[1ex] 
 \hline
\end{tabular}
\caption{Stellar mass, radius, central density and central temperature ratios with respect to the Newtonian values ($M^{N}$, $R^{N}$, and so on) for a few values of the non-commutative parameter $\epsilon$.}\label{tab}
\end{center}
\end{table}

\section{Light elements burning in low-mass stars}

Let us briefly recall the early phase of the low-mass stars' evolution.
During the pre-Main Sequence stage, a proto-star still contracts due to the gravitational attraction. Such an object is luminous but cold, and it follows a Hayashi track on the H-R diagram \cite{hayashi1961stellar}. It is characterized by the almost constant effective temperature. Its further evolution depends on the particular conditions occurring in its core, which can be also rewritten in terms of the stellar mass. That is, the pre-Main Sequence star eventually departs from the Hayashi track due to certain processes. It can happen on the onset of the radiative core development as the luminosity and/or opacity grows - then the star follows the so-called Henyey track. This evolutionary phase is specific for the stars with masses bigger than about $ 0.6 M_\odot$, although it can also depend on a given theory of gravity \cite{Wojnar:2020txr,Guerrero:2021fnz,Gomes:2022sft}. Since modelling of such objects is much more complicated that stars with lower masses, in what follows, we will focus on objects with masses lower than $ 0.6 M_\odot$.

Independently of the further evolution, the pre-Main Sequence stars can already burn lithium when they still follow the Hayashi tracks. It is so because the core temperature needed to ignite lithium is lower than the one for the hydrogen burning. However, when the central temperature and pressure reach a level where hydrogen ignition takes place and the process is stable\footnote{It means that that energy radiated from the photosphere is balanced by energy produced through hydrogen burning in the core.}, the star enters the Main Sequence phase and it is considered as a true star. In what follows, we will focus on such objects, that is, we will calculate the minimal mass for hydrogen burning %as well as we will obtain the lithium abundance in the low-mass stars' atmosphere 
in Snyder model. 

On the other hand, if the object's interior is too cold for hydrogen ignition to begin, the object still contracts till the balance between gravitational contraction and electron degeneracy pressure is not reached. These objects, known as brown dwarfs, lack a source of energy production in their cores and will gradually cool down over time. It is expected that brown dwarf cooling will be also affected by the non-commutative corrections in similar manner as in modified Einstein gravity \cite{Benito:2021ywe,Kozak:2022hdy}.

Having considered the aforementioned properties of the modified Lane-Emden equation, we are now able to determine the minimum main sequence mass (MMSM) in Snyder model. This particular mass represents the threshold mass required for a star to maintain stable thermonuclear reactions within its core and counterbalance energy dissipation on its surface. The magnitudes of thermonuclear reaction rates are predominantly governed by temperature and density, enabling us to estimate the energy generation rate through the application of power laws \cite{Burrows:1992fg}
\begin{equation}\label{Eq: energy generation rate}
\dot{E}_{pp} = \dot{E}_c \left( \frac{T}{T_c}\right) ^s\left( \frac{\rho}{\rho_c}\right) ^{u-1} \ ,
\end{equation}
where the two exponents can be phenomenologically fitted as $ s\approx 6.31 $ and $ u\approx2.28 $ at the transition mass of the core \cite{Burrows:1992fg}, while the function 
\begin{equation}\label{rate}
    \dot{E}_c= \dot{E}_0 \,T_c^s \, \rho_c^{u-1} \, \mathrm{ergs \,g}^{-1}\mathrm{s}^{-1} \ ,
\end{equation}
%la\e_0 esta mal
with $\dot{E}_0 \approx 3.4 \times 10 ^{-9} $ in suitable units. 
For more details, refer to \cite{Burrows:1992fg}. Therefore, in order to calculate the luminosity of the hydrogen burning, we need to integrate (\ref{Eq: energy generation rate}) over the stellar mass:
\begin{equation}\label{Eq: Luminosity}
L_{pp}=\int \dot{E}_{pp}\, dM=  4\pi \dot{\epsilon}_c r_c^3 \rho_c  \int_{0}^{\xi_R} \theta^{\frac{3}{2}\left(u+\frac{2}{3}s\right)}\, \xi^2 d\xi  \ .
\end{equation}
To obtain the second equality, we have used the adiabatic core property that $(T/T_c)=(\rho/ \rho_c)^{2/3}$ and we have changed the integration variables from $M$ to the radial coordinate using (\ref{Eq: star mass}), such that we can then applied the Lane-Emden formalism. Since most of the hydrogen is burnt in the stellar core, we can use the approximation (\ref{Eq: Theta}) in (\ref{Eq: Luminosity}), providing
\begin{equation}\label{Eq: Integral Lpp}
L_{pp}\approx\frac{6 \sqrt{3\pi (1-\epsilon)^3 }}{\omega_{3/2} (2 s+3 u)^{3/2}}\dot{E}_c M \ .
\end{equation}
A typical low-mass star consists of $X=75\%$ of hydrogen\footnote{With $Y=25\%$ of helium, and we will assume that metallicity $Z=0$, however it is very important in a more realistic modelling than in the one we present here.}. Therefore, the number of barions per electron is $\mu_e= \left(X+\frac{Y}{2} \right)^{-1} = 1.143$. Moreover, in such objects the evolution of the electron degeneracy $\eta$ is still important, and although we will not considered here its evolution and dependence on the considered model \cite{auddy2016analytic,Benito:2021ywe,Kozak:2022hdy}, we should take it into account in the polytropic constant
\begin{equation}
    K= \frac{ (3\pi^2)^{2/3} \hbar}{5 m_e m_H^{5/3} \mu_e^{5/3}} \left( 1+ \frac{\alpha_d}{\eta} \right),
\end{equation}
where $m_H$ is the proton mass while $\alpha_d\equiv5\mu_e/2\mu\approx4.82$, where $\mu$ is the mean molecular weight of ionized hydrogen/helium mixtures while $\eta=\Psi^{-1}$. The provided value is for the considered fractions $X$ and $Y$. Then, using it in (\ref{Eq: central dens}) and (\ref{Eq: temperature}) and applying the results to (\ref{rate}), yields the following form of the luminosity (\ref{Eq: Integral Lpp}):
\begin{equation} \label{eq:Lpp}
	L_{pp} =1.54 \times 10^7  L_ \odot \frac{\delta^{5.49} \,(1-\epsilon )^{3/2}}{ \gamma ^{16.46} \omega} M_{-1}^{11.97}  \,\frac{\eta^{10.15}}{ \left(\alpha_d +\eta \right)^{16.46}} \ ,
\end{equation}
where $M_{-1}=M/(0.1M_{\odot})$. This is a partial result of this section. Moreover, the given description can be only applied to the interior region of the star up to a photosphere. Roughly speaking, the photosphere is an outer, luminous layer which one defends as a radius for which the optical depth is (\cite{hansen2012stellar})
 \begin{equation}\label{tau}
     \tau (r) = \int_{r_{ph}}^\infty \kappa_{op} \, \rho \, dr=\frac{2}{3} \ .
 \end{equation}
 The opacity $\kappa_{op}$ describes the optical properties of matter\footnote{ That is, it says how opaque matter is to the electromagnetic radiation.} and it was demonstrated to be a crucial quantity in the stellar modelling. In the further part we will take it as a constant value in order to able to continue the analytical considerations although we should be aware that this is one of the most important parts of our modelling which needs to be improved in the nearest future.
 
Moreover, the photosphere lies close to the surface of the stellar object and its radius, denoted above as $r_\mathrm{ph}$, is well approximated by the star's radius, $R$. Another safe approximation is that the photospheric temperature can be considered as  the effective one, that is, the one which appears in the Stefan-Boltzmann equation
\begin{equation}\label{stef}
    L=4\pi\sigma R^2 T^4_\mathrm{eff} \ .
\end{equation}
The parameter $\sigma$ is the Stefan-Boltzmann constant while $L$ is the luminosity of a black body with temperature $T_\mathrm{eff}$. Therefore, the luminosity of our low-mass object can be also described by the above expression.

Let us note that the optical depth, when $\kappa_{op}=\textrm{constant}$, can be used to integrate the hydrostatic equilibrium equation at the photospheric region. Writing it as
\begin{equation}\label{Eq: hydrostatic eq 2}
p'_{ph} =-\rho g \, ,
\end{equation}
where $g$ is the surface gravity defined as
\begin{equation}\label{Eq: g}
g\equiv \frac{ G M(r)}{ r^2} \approx\textrm{constant},
\end{equation}
and applying to (\ref{tau}) will provide that the photospheric pressure is given simply as
 \begin{equation}\label{fotopress}
     p_{ph}= \frac{2g}{3 \kappa_{op}} \ .
 \end{equation}
The above expression confirms that the opacity is a crucial element for the photosphere's description. In further part, we will consider the Rosseland mean opacity $\kappa_{op}=10^{-2}$cm$^2/$g.
Moreover, we can assume that the photosphere is made of particles whose behaviour is well described by the ideal gas properties. Therefore, the pressure (\ref{fotopress})  can be equalled to
\begin{equation}\label{Eq: ideal gas}
\frac{ \rho _{ph} k_B T_{ph}}{\mu  m_H}=\frac{2 g }{3 \kappa_{op}} \ .
\end{equation}
On the other hand, one finds the photospheric temperature by matching the specific entropies of the gas/metallic phases \cite{Burrows:1992fg}, providing
\begin{equation}\label{Eq: photospheric temperature}
T_{ph}=\frac{1.8 \times 10^6 \rho _{ph}^{0.42}}{\eta ^{1.545}} \ .
\end{equation}
Writing down the surface gravity $ g $ defined (\ref{Eq: g}) as
\begin{equation}\label{Eq: g final}
g=\frac{G^3 M^{5/3}}{\gamma^2 K^2} \ ,
\end{equation}
and applying it, together with photospheric temperature (\ref{Eq: photospheric temperature}), to (\ref{Eq: ideal gas}), one can obtain the photospheric density as a function of mass of a star:
\begin{equation}\label{Eq: photospheric density}
\rho _{ph}=2.957 \times 10^{-5} \frac{\eta ^{1.09} G^{2.11} M^{1.17} \left(\mu  m_H  \right)^{0.70}}{\left(\gamma K\right)^{1.41}(k_B \kappa _{op})^{0.70}} \ .
\end{equation}
To get the photospheric temperature as a function of mass, we can apply the above expression to (\ref{Eq: photospheric temperature}), yielding
\begin{equation}\label{Eq: final photospheric temp}
T_{ph}=2.254 \times 10^4 \frac{G^{0.89} M^{0.49} \left(\mu  m_H \right)^{0.30}}{\eta ^{1.09} \left(\gamma K\right)^{0.59}(k_B \kappa _R)^{0.30}},
\end{equation}
allowing now to write the photospheric luminosity (\ref{stef}) also in terms of the star mass $M$ as
\begin{equation}\label{luminisity}
L_{ph}=	0.534 L_\odot \frac{ M_{-1}^{1.31}}{\eta ^{3.99} \gamma^{0.37} (\alpha_d +\eta )^{0.37} \kappa_{-2}^{1.18}} \ ,
\end{equation}
where $\kappa_{-2}=\kappa_R/(10^{-2} \mathrm{cm}^2\mathrm{g}^{-1})$.

\begin{figure}[t]
\centering
\includegraphics[scale=0.7]{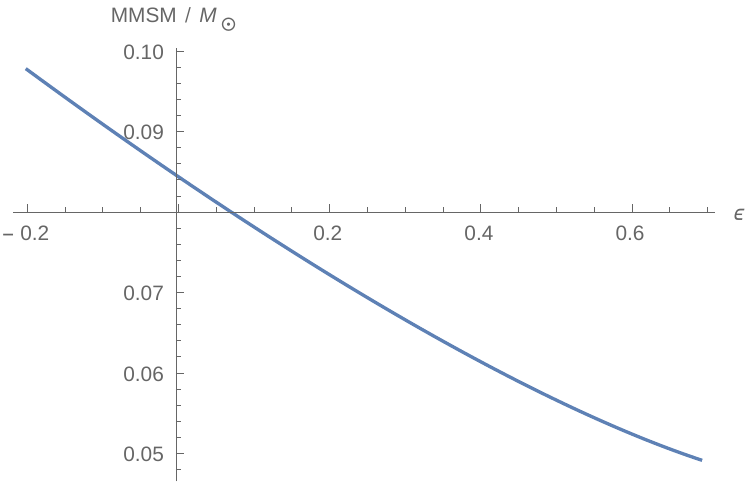}
\caption{The minimum main sequence mass (MMSM) as a function of the parameter $\epsilon$ which carries the information about the Snyder deformation parameters.}
\label{fig1}
\end{figure}

As mentioned, the star burns hydrogen in a stable way if the energy radiated from the photosphere is balanced by energy produced through hydrogen burning in the core. We can write this condition as $L_{ph}=L_{pp}$ given by (\ref{luminisity}) and (\ref{eq:Lpp}), and solve it with respect to the stellar mass. It provides the Minimum Main Sequence Mass (MMSM):
\begin{equation} \label{eq:MMSM}
   M_{-1}^{\mathrm{\tiny MMSM}}=0.227\frac{ \gamma^{1.51} \omega^{0.09} (\alpha_d +\eta )^{1.51} }{(1-\epsilon)^{0.14}\delta^{0.51} \eta ^{1.33}\kappa_{-2}^{0.11}},
\end{equation}
Note that $ (\alpha_d +\eta )^{1.51}/ \eta ^{1.33}$ possesses a unique minimum value for a typical electron degeneracy $\eta=34.7$ for the considered class of astrophysical bodies (we take the Rossland opacity, that is, $\kappa_{-2}=1$). Therefore, if the mass is too low, the above equation has no solution; in other words, $M_{-1}^{\mathrm{\tiny MMSM}}$ is the smallest mass for which (\ref{eq:MMSM}) is satisfied. For Newtonian gravity, the MMSM is $M_\mathrm{\tiny Newt}=0.084M_\odot$. Clearly, altering gravitational or microphysical interactions, we will deal with a different set of solutions. We depict them on the Fig. \ref{fig1}; that is, MMSM as a function of the parameter $\epsilon$.

Taking into account that the lowest observed M dwarf star is
G1 866C \cite{segransan2000accurate} with mass $M = 0.0930 \pm 0.0008M_\odot$, which is obtained for $\epsilon\approx-0.1$, while the values smaller than that are excluded as they provided higher masses. Therefore, $\epsilon\geq-0.1$, which corresponds to (since $\epsilon=\frac{7}{5}\varepsilon\rho_c^\frac{2}{3}=6.2703\rho_c^\frac{2}{3}{\Omega}\times 10^{-52}$)
\begin{equation}
{\Omega} \geq -0.016\times 10^{48}
\end{equation}
for the typical central energy density $\rho_c\sim 10^6$ kg/m$^3$ in these objects.
%, the above bound becomes $ \alpha\omega \geq -0.016\times 10^{48}$.
In $D=3$, we therefore have:
%$\Omega =\beta\left( 4\chi -\frac{3}{2}\right)   $
\begin{equation}
  {\Omega= \beta\left( 4\chi -\frac{3}{2}\right) }\geq -1.6\times 10^{46}  
\end{equation}
and this gives two possibities depending on realization choice: 
\begin{itemize}
    \item for $\left( 4\chi -\frac{3}{2}\right) >0$, implying $\chi>\frac{3}{8}$, we get:
$$    \beta\geq -\frac{1.6}{(4\chi-\frac{3}{2})} \times 10^{46},$$
which is obvious as we assumed $\beta>0$, 
\footnote{It is worth to point out that one can consider models called anti-Snyder, see e.g. \cite{Mignemi:2011gr} with negative $\beta$. Such possibility of negative parameter in GUP theories was also investigated, see e.g. \cite{Ong:2018zqn,Buoninfante:2019fwr}} 
\item for $(4\chi-\frac{3}{2})<0$, implying $\chi<\frac{3}{8}$, we get:
$$    \beta\leq \frac{1.6}{(\frac{3}{2}-4\chi)} \times 10^{46}.$$
For example, choosing the value of $\chi=0$ (which was considered in e.g. \cite{Maggiore:1993rv},\cite{Maggiore:1993zu} and studied in many GUP related effects) we get: 
$$    \beta\leq 1.067 \times 10^{46}$$
or, taking the dimensionless parameter, our bound takes 
\begin{equation}\label{beta0}
   \beta_0 \leq 4.5 \times 10^{47}, 
\end{equation}
where we have used a more common notation, that is, $\beta_0=\beta M^2_Pc^2$. It seems that our investigation provides a better upper bound based on astrophysical effects than obtained in literature so far, up to our knowledge, see e.g. \cite{Bosso:2023aht}. The best astrophysical constraint was obtained by examining the perihelion precession in the Solar System \cite{Scardigli:2014qka}, that is, $\beta\leq10^{69}$. However, it is worth to point out that better bounds were obtained based on the microscopic effects (for review, see \cite{Bosso:2023aht}), for example in \cite{Wagner:2023fmb} it was shown that $\beta<10^{16}$. Although our stellar modelling is very simple, it should be underlined that it is crucial to take into account more realistic matter description in stellar objects to obtain better bounds.
Note that considering values of $\chi<0$ can lower the bound; for example choosing $\chi=-20$ we improve the bound by 2 orders of magnitude. Let us recall that $\chi$ parametrizes the choice of the realization for the Snyder model, appearing in the phase-space relation (\ref{gen_real_p-x}) and gives us certain freedom to assign a numerical value.
\end{itemize}

\section{Conclusions}

The aim of this paper was to investigate how quantum gravitational corrections affect stellar interiors of low-mass stars.
Considering NC space-time, with an example of the Snyder model, allowed us to check how various parametrizations of the deformed phase space, stemming from the choice of representation of the Snyder space, affect physical solutions. 
We have relied on the Fermi equation of state with the non-commutative (quantum-gravitational) corrections which we previously derived in \cite{Pachol:2023tqa}, with the change of the measure to (\ref{measure}). Applying such modified polytrope to the Poisson equation, together with the hydrostatric equilibrium one, allowed us to show equivalence to a modified Lane-Emden equation which is used to determine properties of convective stars. Considering the toy-model assumptions, one obtains that generally negative values of the parameter $\epsilon$ make the star bigger and more massive, however with lower central densities and temperatures, making the stellar material more incompressible, as discussed in our previous work \cite{Pachol:2023tqa}. On the other hand, positive values enhance the compressibility of the stellar material, at the same time increasing the central parameters with decrease in stellar radius and mass.
We have also pointed out that such modifications of the Lane-Emden equation occur also, and are indistinguishable from theoretical point of view, in modified Einstein gravity. Therefore, it can be treated as a Lane-Emden equation resulting from some quantum gravitational proposal which is related to the modified Poisson equation of the following form
\begin{equation}
   \nabla^2\phi =  4\pi G \rho - \tilde\epsilon \nabla^2 \rho^\frac{4}{3}.
\end{equation}
This study shows that, for example NC geometry or modified Einstein gravity, both providing quantum gravitational corrections to classical solutions, can serve as an effective description of the theory without full knowledge of quantum gravity itself and can guide us in a search of observable and measurable effects related to quantum gravity whatever it may be.

Nevertheless, Lane-Emden formalism is also useful when one wants to incorporate more realistic physics and to obtain predictions of a particular processes from a given model. One of those, which we have studied in the present manuscript, is the energy generation rate resulting from the thermonuclear reactions which happen in the stellar cores. Here, we have focused on the hydrogen ignition which is a crucial phenomena in the stellar evolution - when this light element starts burning, the so-far contracting object becomes "a true" star, that is, it enters the Main Sequence phase. At the same time, the contraction stops as the gravitational pull us balanced by the pressure which is a result of the energy transport from the stellar core to its surface.

The mentioned process depends on any modifications introduced to the equation of state or/and quantum gravity. It is so because the energy generation rates strongly depend on core's properties which are sensitive to the quantum and gravitational interactions. Because of that fact, we deal with an opportunity to constrain the parameters of the Snyder model with astrophysical observations. As shown, the NC corrections also alter the so-called Minimum Main Sequence Mass which is a critical mass which a star needs to have in order to be considered as a true star (a star burning hydrogen in a stable way). The lowest mass of a true star is known to be $M = 0.0930 \pm 0.0008M_\odot$, so any value of the minimal length parameter, arising from the Snyder model, making the minimum mass bigger than this, should be ruled out. This allows to put the bound
\begin{equation*}
\beta_0 \leq 4.5 \times 10^{47}
\end{equation*}
for Snyder model realizations with $\chi<0.375$, 
in the case of a typical stellar object studied in this paper. Definitively, our bound is better than the ones obtained in astrophysical frameworks studied in other papers (see the table 2 in \cite{Bosso:2023aht}), but poor when compared to the tabletop experiments not related to gravity (table 1 therein). However, our bound could be still improved when more sophisticated description of matter inside a stellar object would be taken into account. Note that our modelling should still be treated as a toy-model. We have simplified a lot the photosphere description (ideal gas, no phase transition between the interior and photospheric regions, no ionization processes, poor model for the opacity and no modification to the gravitational counterpart of photospheric properties) as well as the interior region - the main improvement would be to take into account the time-dependence of the electron degeneracy which was shown to have a non-trivial effect on the MMSM and evolution of this class of objects \cite{auddy2016analytic,Benito:2021ywe,Kozak:2022hdy}. Secondly, we have used only one object to constrain the model while the proper statistical and uncertainties analysis should be also carried out, providing the confidence level. 
Additionally, in our approach, considering the most general realization of the Snyder model and the deformed phase-space associated with it, leading to the GUP, we have a freedom provided by the choice of representations of Snyder model (parameter $\chi$) and this  allows us to obtain even better bounds.

Nevertheless, the most important result of this paper is demonstrating that taking into account more realistic description of matter than usually undertaken by the GUP community in regards to the astrophysical objects, allowed us to obtain much more stringent astrophysical bound.

\section*{Acknowledgements}

The authors would like to acknowledge networking support by the COST Action CA18108 and STSM Grant No. E-COST-GRANT-CA18108-21665909. AP has been supported by the Polish National Science Center (NCN), project 2022/45/B/ST2/01067. AW acknowledges financial support from MICINN (Spain) {\it Ayuda Juan de la Cierva - incorporac\'ion} 2020.

\section*{Appendix: details on the measure}
To obtain the deformed phase space measure (\ref{measure}) based on the Liouville theorem we have relied on the results obtained in \cite{Chang:2001bm} with the following identification:  $\beta \rightarrow \beta \left( \chi-\frac{1}{2}\right) $ and 
$\beta ^{\prime }\rightarrow 2\chi\beta $. Therefore the deformation of the phase space corresponding to (\ref{gen_real_p-x}) and using eq. (9) from \cite{Chang:2001bm} in $D=3$ gives: 
\begin{equation}\label{mes1}
 \frac{d^3xd^{3}p}{\left( 1+\beta \left( \chi -\frac{1}{2}\right)
p^{2}\right) ^{2}\left( 1+\beta \left( 3\chi -\frac{1}{2}\right)
p^{2}\right) ^{\sigma }},
\end{equation}
where $\sigma = \frac{4\chi -1}{6\chi -1}$.
Since we are interested in the results up to the first order in the non-commutativity parameter $\beta$, as the phase space (\ref{gen_real_p-x}) is described to that order only, we note that the above expression reduces to:
\begin{eqnarray}
d^3xd^3p \left[ 1-\Omega p^{2}+O(p^4)\right]\label{expansion}
\end{eqnarray}
where the $\Omega$ coefficient appearing in this expansion, linear in $\beta$, is:
\footnotesize{
\begin{equation}\label{nowe_ao}
 \Omega =2\beta \left( \chi -\frac{1}{2}\right)
+\beta \frac{4\chi -1}{6\chi -1}\left( 3\chi -\frac{1}{2}\right) =\beta
\left( 4\chi -\frac{3}{2}\right)   .
\end{equation}} \normalsize
On the other hand, up to the linear order in $\beta$, we have the following equality:
\begin{equation}
    \frac{1}{1+\Omega p^{2}}=1-\Omega p^{2}+O(p^4)
\end{equation}
Therefore, (\ref{mes1}) can be simply rewritten as $ \frac{1}{1+\Omega p^{2}}$, if considered up to linear order in $\beta$.

Now we recall the next steps \cite{Pachol:2023tqa}, the partition function in the {grand-canonical ensemble} is given as 
\begin{equation}\label{part1}
    \mathrm{ln}Z = \sum_i \mathrm{ln}\left[1+az e^{-E_i/k_BT}\right]
\end{equation}
where $T$ is the temperature, $k_B$ Boltzmann constant, $z=e^{\mu/k_BT}$ while $\mu$ is the chemical potential and $a=1$ ($a=-1$) if the particles are fermions (bosons). 

If we consider a large volume, the summation in the above partition function (\ref{part1}) should be replaced by
\begin{equation}
      \sum_i\rightarrow \frac{1}{(2\pi \hbar)^3} \int \frac{d^3xd^3p}{(1+\Omega p^2)} 
\end{equation}
and in the spherical symmetric case we obtain
\begin{equation}\label{partitionCORR}
     \mathrm{ln}Z =  \frac{4\pi V}{(2\pi \hbar)^3}\frac{g}{a}\int \mathrm{ln}\left[1+az e^{-E/k_BT}\right] \frac{p^2 dp}{1+\Omega p^2} \ 
\end{equation}
where we took $V=\int d^3x$. Considering pressure given by
\begin{equation}
    P= k_B T \frac{\partial}{\partial V}  \mathrm{ln}Z 
\end{equation}
and integrating by parts\footnote{We take $\int u(p)v'(p) dp=uv-\int u'(p)v(p)$ with $u(E(p))=\ln(h(E(p)))$, where $h(E(p))=1+az e^{-E/k_BT}$ and $E=\sqrt{p^2c^2+m^2c^4}$, and $v'(p)=\frac{p^2}{1+\Omega p^2}$. We calculate $u'=\frac{h'(E)}{h(E)}\frac{d}{dp}E(p)$ and $v=\int \frac{p^2}{1+\Omega p^2} dp=\frac{1}{3}p^3\, _{2}F_{1}\left(\frac{3}{2},{1},\frac{5}{2},-p^2 \Omega\right)=
\frac{p}{\Omega}-\frac{\mathrm{ArcTan}[p \sqrt{\Omega}}{\Omega^{3/2}}$.} the above expression we obtain a special case of the hypergeometric function (see \cite{pachol2023fermi} and take $\alpha=1$, compare it also to \cite{Tunacao:2022ffq}):
\begin{align}\label{result0}
     P&=  \frac{1}{\pi^2 \hbar^3}\int \frac{1}{3}p^3\, _{2}F_{1}\left(\frac{3}{2},{1},\frac{5}{2},-p^2 \Omega\right) f(E) \frac{c^2 p}{E}dp  \\ \nonumber
     &=  \frac{1}{\pi^2 \hbar^3}\int \left( \frac{p}{\Omega} - \frac{\text{ArcTan}[p\sqrt{\Omega}]}{\Omega^\frac{3}{2}}
     \right)
     f(E) \frac{c^2 p}{E}dp .
\end{align}

For the case when $|\Omega p^2| << 1$ for the hypergeometric function
\footnote{{Note that this condition is satisfied thanks to the fact that $\beta$ is small and it is not a restriction. Nevertheless, we can consider two cases when $\Omega>0$ leading to $\chi>0.375$ and $\Omega <0$ leading to $\chi<0.375$.}}, we can write the pressure as
\begin{equation}
      P=  \frac{1}{\pi^2 \hbar^3}\int \frac{p^3}{3} \left( 
      \sum_{k=0}^{\infty} \frac{\left(\frac{3}{2}\right)_k(-{\Omega p^2})^k}{ \left(\frac{5}{2}\right)_k k!}
      \right) f(E) \frac{c^2 p}{E}dp .
\end{equation}
Taking into account the terms up to the first order in $\beta$ (i.e. $\Omega$), as we consider the NC deformation only up to linear terms in $\beta$ cf. (\ref{gen_real_p-x}), then the first two terms of the series we have
\begin{equation}\label{1order}
      P=  \frac{1}{\pi^2 \hbar^3}\int \left(\frac{p^3}{3} -\frac{{\Omega} p^5}{5} \right) f(E) \frac{c^2 p}{E}dp \ .
\end{equation}
%We note that in our approach {$\Omega$} depends on the realization of the deformation of the phase space $\chi$, hence we can consider cases when $\Omega=0$, $\Omega>0$ or $\Omega<0$. We discuss the implications of this in detail in Sec. 4. 

\bibliographystyle{spphys} 
\bibliography{biblio.bib}

\end{document}